\documentclass[a4paper,11pt]{article}
\pdfoutput=1 

\usepackage{jcappub} 

\usepackage[T1]{fontenc} 
\usepackage{ifthen}
\usepackage{epsfig}
\usepackage{booktabs} 
\usepackage{natbib}
\usepackage{bbold}
\usepackage{diagbox}

\newcommand{\beqra}{\begin{flalign}}
\newcommand{\eeqra}{\end{flalign}}
\newcommand{\beq}{\begin{equation}}
\newcommand{\eeq}{\end{equation}}

\newcommand{\onehalf}{\tfrac{1}{2}}

\newcommand{\cost}{\cos\theta}

\newcommand{\ERmax}{E_R^{\textrm{max}}}
\newcommand{\ERmin}{E_R^{\textrm{min}}}

\newcommand{\ELi}{E_L^{\,i}}
\newcommand{\EHi}{E_H^{\,i}}
\newcommand{\OmCap}{\Omega_v^-(w)}

\newcommand{\hO}{\hat{O}}

\newcommand{\vv}{\mathbf{v}}

\newcommand{\vw}{\mathbf{w}}

\usepackage{ccicons} 
\usepackage{listings}
\usepackage{color}   
\usepackage{braket}  
\usepackage{hyperref}
\usepackage{dsfont}  
\usepackage{leftidx} 
\RequirePackage{mathrsfs}
\RequirePackage{dsfont}
\usepackage{bm}
\usepackage{placeins}
\usepackage{float}
\allowdisplaybreaks

\title{\boldmath New constraints on inelastic dark matter from IceCube}
\author[a]{Riccardo Catena}
\author[b,c]{and Fredrik Hellstr\"om}
\affiliation[a]{Chalmers University of Technology, Department of Physics, SE-412 96 G\"oteborg, Sweden}
\affiliation[b]{Chalmers University of Technology, Department of Electrical Engineering, SE-412 96 G\"oteborg, Sweden}
\affiliation[c]{University of Gothenburg, Department of Physics, SE-412 96, Gothenburg, Sweden}
\emailAdd{catena@chalmers.se}
\emailAdd{frehells@chalmers.se}

\abstract{We study the capture and subsequent annihilation of inelastic dark matter (DM) in the Sun, placing constraints on the DM-nucleon scattering cross section from the null result of the IceCube neutrino telescope.~We then compare such constraints with exclusion limits on the same cross section that we derive from XENON1T, PICO and CRESST results.~We calculate the cross section for inelastic DM-nucleon scattering within an extension of the effective theory of DM-nucleon interactions which applies to the case of inelastic DM models characterised by a mass splitting between the incoming and outgoing DM particle.~We find that for values of the mass splitting parameter larger than about 200 keV, neutrino telescopes place limits on the DM-nucleon scattering cross section which are stronger than the ones from current DM direct detection experiments.~The exact mass splitting value for which this occurs depends on whether DM thermalises in the Sun or not.~This result applies to all DM-nucleon interactions that generate DM-nucleus scattering cross sections which are independent of the nuclear spin, including the ``canonical'' spin-independent interaction.~We explicitly perform our calculations for a DM candidate with mass of 1 TeV, but our conclusions qualitatively also apply to different masses.~Furthermore, we find that exclusion limits from IceCube on the coupling constants of this family of spin-independent interactions are more stringent than the ones from a (hypothetical) reanalysis of XENON1T data based on an extended signal region in nuclear recoil energy.~Our results should be taken into account in global analyses of inelastic DM models.}

\begin{document}
\maketitle
\flushbottom

\section{Introduction}
\label{sec:introduction}
The existence of invisible mass, or dark matter (DM), in our Universe is supported by observations performed on very different physical scales.~These include the anomalous motion of stars and galaxies, gravitational lensing events in cluster of galaxies, patterns in the anisotropies of the cosmic microwave background radiation, and the observed large scale structure of the Universe (see~\cite{Bertone:2004pz} and references therein for a comprehensive review).~In the leading paradigm of modern cosmology DM is made of hypothetical particles with interactions at the weak scale or below~\cite{Arcadi:2017kky}.~This class of DM particles is currently searched for using, e.g., direct detection experiments~\cite{Goodman:1984dc}, which look for nuclear recoils induced by the non-relativistic scattering of DM particles in low-background detectors, and indirect detection experiments, which search for DM annihilation signals produced in space or at the centre of the Sun or of the Earth~\cite{Krauss:1985ks,Freese:1985qw}, where DM is expected to accumulate by losing energy while scattering off nuclei in the solar and terrestrial interiors.~Among the indirect detection experiments, neutrino telescopes, such as IceCube~\cite{Aartsen:2016zhm}, which search for neutrinos from DM annihilation in the Sun are of special interest for this work.

Interpreting the null result of current experiments, DM is commonly assumed to scatter off nuclei elastically, i.e.~the DM particle is in the same state before and after scattering, e.g.~\cite{Lewin:1995rx}.~While this assumption is often fulfilled by popular models for DM~\cite{Bergstrom00}, it is not always true.~For example, in DM-nucleus collisions a DM particle could scatter to an excited state of higher mass (endothermic reaction), or scatter from an excited state to a different state of lower mass (exothermic reaction)~\cite{TuckerSmith:2001hy}\footnote{An alternative scenario is the one where the target nucleus is scattered off to an excited state~\cite{Goodman:1984dc}.~This scenario will not be discussed here.}.~The family of models where DM scatters off nuclei inelastically is collectively referred to as inelastic DM.~Inelastic DM has initially been proposed as an attempt to reconcile the observation of an annual modulation in the rate of nuclear recoil events recorded by the DAMA collaboration with the null result reported by other experiments~\cite{TuckerSmith:2001hy}.~In this context, it has also been noticed that inelastic DM-nucleus scattering can occur in a variety of theories, including supersymmetric models of nearly pure Higgsinos~\cite{Fox:2014moa}, magnetic inelastic DM~\cite{Chang:2010en}, and dark photon mediated DM~\cite{Smolinsky:2017fvb}.~While the initial motivation based on reconciling DAMA with the null result from other experiments has become less attractive due the strong exclusion limits presented by the LUX, XENON and PandaX collaborations, the fact that inelastic DM appears naturally in a variety of frameworks holds true.~Furthermore, it has been shown that within specific realisations of inelastic DM, the range of mass splittings between incoming and outgoing DM particles can be broader than initially proposed~\cite{Bramante:2016rdh}.~The large mass splitting limit of inelastic DM is known as the ``inelastic frontier''.

The kinematics of inelastic DM-nucleus scattering is significantly different from the one of elastic interactions~\cite{Menon:2009qj,Bozorgnia:2013hsa,Scopel:2014kba,Scopel:2015eoh,Blennow:2015hzp,DelNobile:2015lxa}.~In particular, for DM particles heavier than atomic nuclei, e.g.~of mass 1 TeV, the inelastic DM-nucleus scattering is characterised by:~1) A finite minimum velocity the DM particle must have for the scattering to be kinematically allowed which scales like the inverse of the square root of the target nucleus mass;~2) A minimum nuclear recoil energy required for the scattering to occur which is approximately equal to the mass splitting between the incoming and outgoing DM particle.~These properties imply that inelastic DM-nucleus scattering is kinematically favoured for target nuclei with large mass numbers, and that only direct detection experiments which record nuclear recoil energies larger than the inelastic DM mass splitting parameter can effectively probe this scenario.~Based on these properties, it has been found in~\cite{Bramante:2016rdh} that an experiment like CRESST, which probes a range of nuclear recoil energies larger than, e.g., XENON1T, is effective in setting limits on the DM-nucleon scattering cross-section in the large mass splitting limit.~On the other hand, as far as neutrino telescopes are concerned, the inelastic frontier of DM models remains as of yet unexplored.

In this article we set constraints on the DM-nucleon scattering cross section of inelastic DM models in the large mass splitting limit using data from neutrino telescopes~\cite{Aartsen:2016zhm} and direct detection experiments~\cite{Aprile:2018dbl,Amole:2015pla,Angloher:2015ewa}.~The DM-nucleon scattering cross section is computed within the non-relativistic effective theory of DM-nucleon interactions, formulated in~\cite{Fitzpatrick:2012ix}, applied to the analysis of neutrino telescope data and DM capture in the Sun and Earth in~\cite{Liang:2013dsa,Catena:2015uha,Catena:2015iea,Catena:2016ckl,Catena:2016kro,Kavanagh:2016pyr,Widmark:2017yvd}, and extended to inelastic DM in~\cite{Barello:2014uda}.~Our constraints from neutrino telescopes are compared with those we obtain from an analysis of XENON1T, PICO and CRESST results.~We find that in the inelastic frontier, exclusion limits from neutrino telescopes can be stronger than those from direct detection, even for canonical spin-independent DM-nucleon interactions.~This result should be taken into account in the analysis of IceCube data within inelastic DM models.

This article is organised as follows.~In Sec.~\ref{sec:inelastic} we introduce the theory of inelastic DM, focusing on kinematical aspects (Sec.~\ref{sec:kinematics}), and on the expected signals at direct detection experiments (Sec.~\ref{sec:dd}) and neutrino telescopes (Sec.~\ref{sec:nt}).~Sec.~\ref{sec:results} focuses on our limits from neutrino telescopes and direct detection experiments on the DM-nucleon scattering cross section of inelastic DM models in the large mass splitting limit.~We comment on our results and conclude in Sec.~\ref{sec:conclusion}.

\section{Inelastic dark matter}
\label{sec:inelastic}
In this section we review the kinematics of DM-nucleus scattering in inelastic DM models characterised by two DM particle states of different mass (Sec.~\ref{sec:kinematics}).~In the same framework, we also review the theory of DM direct detection (Sec.~\ref{sec:dd}), and the process of DM capture in the Sun (Sec.~\ref{sec:nt}).

\subsection{Kinematics}
\label{sec:kinematics}
We are interested in models with two DM particle states, denoted here by $\chi$ and $\chi^*$.~The two states have masses $m_{\chi}$ and $m_{\chi^*}$, respectively, differing by a mass splitting $m_{\chi^*}-m_{\chi}  \equiv \delta \ll m_\chi$.~In the nucleus rest frame, the energy and momentum conservation equations that govern the non-relativistic scattering process $\chi N \rightarrow \chi^* N$, where $N$ is a nuclear state, take the following form
\begin{align}
\onehalf m_\chi w_i^2 &= \onehalf m_Tv_{N}^2+\onehalf m_{\chi^*}w_f^2+\delta, \label{eq:EconsScat}\\
m_\chi\vw_i&=m_T\vv_{N}+m_{\chi^*}\vw_f,
\end{align}
where $m_T$ is the target nucleus mass, $\vv_{N}$ its final velocity, and $\vw_i$ ($\vw_f$) the initial (final) DM velocity.~Squaring the momentum conservation equation, and denoting by $\theta$ the angle between $\vv_N$ and $\vw_i$, we get
\begin{align}
\onehalf m_{\chi^*}w_f^2 = \frac{m_\chi^2w_i^2+m_T^2v_N^2-2m_Tm_\chi w_iv_N\cost}{2m_{\chi^*}}\,.
\end{align}
By replacing this relation into Eq.~(\ref{eq:EconsScat}), we find the following equation for the nuclear recoil energy $E_R=\onehalf m_Tv_N^2$:
\begin{align}
E_R\left(\frac{m_T}{m_{\chi^*}}+1\right)-\sqrt{2m_T E_R}\frac{m_\chi}{m_{\chi^*}}w_i\cost +\delta-\onehalf m_\chi w_i^2 \frac{\delta}{m_{\chi^*}}=0.\label{eq:ER1stdeg}
\end{align}
Squaring Eq.~(\ref{eq:ER1stdeg}) gives
\begin{align}
E_R^2+E_R\frac{\mu_*}{m_T}\left(2\delta -\frac{m_\chi}{m_{\chi^*}}\delta w_i^2 -2\frac{m_\chi^2}{m_{\chi^*}^2}\cos^2\hspace{-0.07 cm}\theta\,\mu_*w_i^2 \right)&
+\delta^2\frac{\mu_*^2}{m_T^2}\left(1-\frac{m_\chi}{m_{\chi^*}}w_i^2 +\frac{m_\chi^2}{m_{\chi^*}^2}\frac{w_i^4}{4} \right)&=0,
\label{eq:ER2}
\end{align}
where $\mu_*=m_Tm_{\chi^*}/(m_T+m_{\chi^*})$ is the $\chi^*$-nucleus reduced mass.~In the non-relativistic limit, and considering only terms which are at most quadratic in $\delta/m_\chi$ and $w_i$, 
Eq.~(\ref{eq:ER2}) can be simplified as follows
\begin{align}
E_R^2+E_R\frac{\mu}{m_T}\left(2\delta-2\mu w_i^2\cos^2\hspace{-0.07 cm}\theta \right)+\delta^2\frac{\mu^2}{m_T}=0,
\end{align}
where, in analogy with $\mu^*$, $\mu=m_Tm_\chi/(m_T+m_\chi)$.~This equation has maximum and minimum solutions for $E_R$ given by
\begin{align}
\ERmax &= \frac{\mu^2}{m_T}w_i^2\left(1+\sqrt{1-\frac{2\delta}{\mu w_i^2}}\right)-\frac{\mu}{m_T}\delta \,,\label{eq:Emax}\\
\ERmin &= \frac{\mu^2}{m_T}w_i^2\left(1-\sqrt{1-\frac{2\delta}{\mu w_i^2}}\right)-\frac{\mu}{m_T}\delta \,.\label{eq:Emin}
\end{align}
These solutions are real only for
\begin{align}
w_i \geq 
\Re{\sqrt{\frac{2\delta}{\mu}}}\,,\label{eq:wmink}
\end{align}
which is equivalent to requiring that the initial $\chi$-nucleus center of mass energy is larger than the mass splitting $\delta$.~Eq.~(\ref{eq:wmink}) gives the lowest possible DM speed for the scattering to be kinematically allowed.

\begin{table}[t]
    \centering
    \begin{tabular*}{\columnwidth}{@{\extracolsep{\fill}}llll@{}}
    \hline
        $\hat{\mathcal{O}}_1 = \mathds{1}_{\chi}\mathds{1}_N$  & $\hat{\mathcal{O}}_{10} = i{\bf{\hat{S}}}_N\cdot\frac{{\bf{\hat{q}}}}{m_N}\mathds{1}_\chi$   \\
        $\hat{\mathcal{O}}_3 = i{\bf{\hat{S}}}_N\cdot\left(\frac{{\bf{\hat{q}}}}{m_N}\times{\bf{\hat{v}}}^{\perp}\right)\mathds{1}_\chi$ & $\hat{\mathcal{O}}_{11} = i{\bf{\hat{S}}}_\chi\cdot\frac{{\bf{\hat{q}}}}{m_N}\mathds{1}_N$   \\
        $\hat{\mathcal{O}}_4 = {\bf{\hat{S}}}_{\chi}\cdot {\bf{\hat{S}}}_{N}$ & $\hat{\mathcal{O}}_{12} = {\bf{\hat{S}}}_{\chi}\cdot \left({\bf{\hat{S}}}_{N} \times{\bf{\hat{v}}}^{\perp} \right)$  \\
        $\hat{\mathcal{O}}_5 = i{\bf{\hat{S}}}_\chi\cdot\left(\frac{{\bf{\hat{q}}}}{m_N}\times{\bf{\hat{v}}}^{\perp}\right)\mathds{1}_N$ & $\hat{\mathcal{O}}_{13} =i \left({\bf{\hat{S}}}_{\chi}\cdot {\bf{\hat{v}}}^{\perp}\right)\left({\bf{\hat{S}}}_{N}\cdot \frac{{\bf{\hat{q}}}}{m_N}\right)$\\
        $\hat{\mathcal{O}}_6 = \left({\bf{\hat{S}}}_\chi\cdot\frac{{\bf{\hat{q}}}}{m_N}\right) \left({\bf{\hat{S}}}_N\cdot\frac{\hat{{\bf{q}}}}{m_N}\right)$ &  $\hat{\mathcal{O}}_{14} = i\left({\bf{\hat{S}}}_{\chi}\cdot \frac{{\bf{\hat{q}}}}{m_N}\right)\left({\bf{\hat{S}}}_{N}\cdot {\bf{\hat{v}}}^{\perp}\right)$  \\
        $\hat{\mathcal{O}}_7 = {\bf{\hat{S}}}_{N}\cdot {\bf{\hat{v}}}^{\perp}\mathds{1}_\chi$ &  $\hat{\mathcal{O}}_{15} = -\left({\bf{\hat{S}}}_{\chi}\cdot \frac{{\bf{\hat{q}}}}{m_N}\right)\left[ \left({\bf{\hat{S}}}_{N}\times {\bf{\hat{v}}}^{\perp} \right) \cdot \frac{{\bf{\hat{q}}}}{m_N}\right] $ \\
        $\hat{\mathcal{O}}_8 = {\bf{\hat{S}}}_{\chi}\cdot {\bf{\hat{v}}}^{\perp}\mathds{1}_N$ &  $\hat{\mathcal{O}}_{17}=i \frac{{\bf{\hat{q}}}}{m_N} \cdot \mathbf{\mathcal{S}} \cdot {\bf{\hat{v}}}^{\perp} \mathds{1}_N$ \\
        $\hat{\mathcal{O}}_9 = i{\bf{\hat{S}}}_\chi\cdot\left({\bf{\hat{S}}}_N\times\frac{{\bf{\hat{q}}}}{m_N}\right)$ & $\hat{\mathcal{O}}_{18}=i \frac{{\bf{\hat{q}}}}{m_N} \cdot \mathbf{\mathcal{S}}  \cdot {\bf{\hat{S}}}_{N}$  \\
    \hline
    \end{tabular*}
    \caption{Quantum mechanical operators defining the non-relativistic effective theory of DM-nucleon interactions~\cite{Fitzpatrick:2012ix}.~The operators are expressed in terms of the basic invariants under Galilean transformations:~the momentum transfer, $\mathbf{\hat{q}}$, the transverse relative velocity operator $\mathbf{\hat{v}}^\perp$, the nucleon and DM spin operators, denoted by $\mathbf{\hat{S}}_N$ and $\mathbf{\hat{S}}_\chi$, respectively, and the identities in the nucleon and DM spin spaces, $\mathds{1}_{\chi}$ and $\mathds{1}_N$.~All operators have the same mass dimension, and $m_N$ is the nucleon mass.~In the case of inelastic DM, $\mathbf{\hat{v}}^\perp$ and the DM-nucleus relative velocity operator, $\mathbf{\hat{v}}$, are related by the equation $\mathbf{\hat{v}}^\perp=\mathbf{\hat{v}}+\mathbf{\hat{q}}/(2\mu_N)+\delta \, \mathbf{\hat{q}}/q^2$, where $\mu_N$ is the DM-nucleon reduced mass and $q$ the momentum transfer~\cite{Barello:2014uda}.~Standard spin-independent and spin-dependent interactions correspond to the operators $\hat{\mathcal{O}}_{1}$ and $\hat{\mathcal{O}}_{4}$, repsectively, while $\mathbf{\mathcal{S}}$ is a symmetric combination of spin 1 polarisation vectors~\cite{Dent:2015zpa}.~The operators $\hat{\mathcal{O}}_{17}$ and $\hat{\mathcal{O}}_{18}$ can only arise for spin 1 DM.~Following~\cite{Fitzpatrick:2012ix}, here we do not consider the interaction operators $\hat{\mathcal{O}}_{2}$ and $\hat{\mathcal{O}}_{16}$:~the former is quadratic in $\mathbf{\hat{v}}^\perp$ (and the effective theory expansion in~\cite{Fitzpatrick:2012ix} is truncated at linear order in $\mathbf{\hat{v}}^\perp$ and second order in $\mathbf{\hat{q}}$) and the latter is a linear combination of $\hat{\mathcal{O}}_{12}$ and $\hat{\mathcal{O}}_{15}$.}
\label{tab:operators}
\end{table}

\subsection{Direct detection}
\label{sec:dd}
The differential rate of nuclear recoil events per unit detector mass in a DM direct detection experiment is given by
\begin{align}
\frac{{\rm d} R}{{\rm d} E_R} = \sum_T \xi_T \frac{\rho_\chi}{m_\chi m_T} \int_{|\vw| \ge w_{\rm min}} {\rm d}^3 v \, |\vw| f(\vw) \, \frac{{\rm d}\sigma_T}{{\rm d} E_R} (w^2, E_R) \,,
\label{eq:rate}
\end{align}
where $\vw\equiv \vw_i$, and $w_{\rm min}$ is the minimum kinematically allowed DM speed for a given nuclear recoil energy $E_R$.~In Eq.~(\ref{eq:rate}), $\rho_\chi$ is the local DM density, $f(\vw)$ is the DM velocity distribution in the detector rest frame, and the sum runs over all elements in the detector, each giving a contribution weighted by the corresponding mass fraction $\xi_T$.~We calculate the differential cross section for DM-nucleus scattering in Eq.~(\ref{eq:rate}), ${\rm d}\sigma_T/{\rm d} E_R$, within the non-relativistic effective theory of DM-nucleon interactions~\cite{Fitzpatrick:2012ix}.~The theory is characterised by 16 DM-nucleon interaction operators, labelled by an index $j$ and listed in Tab.~\ref{tab:operators}, and 8 nuclear response functions describing the response of nuclei to the interactions in Tab.~\ref{tab:operators}~\cite{Anand:2013yka,Catena:2015uha}.~In general, ${\rm d}\sigma_T/{\rm d} E_R$ depends on the velocity $w$, on the nuclear recoil energy $E_R$, on the mass splitting parameter $\delta$, on the DM mass, on the DM and nuclear spins, and on isoscalar and isovector coupling constants, $c_j^0$ and $c_j^1$, respectively.~For further details, and an explicit expression for ${\rm d}\sigma_T/{\rm d} E_R$ which applies to the case of inelastic DM, see~\cite{Anand:2013yka}.~For the local DM density, we assume the value $\rho_\chi=0.4$~GeV~cm$^{-3}$, e.g.~\cite{Catena:2009mf}.~For the DM velocity distribution in the detector rest frame, we adopt a Maxwellian distribution with a Galactic escape velocity of $544$~km~s$^{-1}$ and a circular speed of $220$~km~s$^{-1}$ for the local standard of rest (i.e.~the so-called Standard Halo Model~\cite{Freese:2012xd}).

In the case of inelastic DM, the minimum kinematically allowed DM speed, $w_{\rm min}$, can be found from Eq.~(\ref{eq:ER1stdeg}), which in the non-relativistic limit, and assuming $\delta/m_\chi \ll 1$, reads as follows 
\begin{align}
E_R \frac{m_T}{\mu} - w_i\cost\sqrt{2m_TE_R}+\delta=0\,.
\end{align}
For a given $E_R$, the minimum speed is given by
\begin{align}
w_{\rm min} = \frac{E_R\tfrac{m_T}{\mu}+\delta}{\sqrt{2m_TE_R}},
\label{eq:wmin}
\end{align}
which, as a function of $E_R$, has an absolute minimum at
\begin{align}
E_R^{\rm min} = \frac{\mu}{m_T}\delta
\label{eq:ERmin}
\end{align}
given by Eq.~(\ref{eq:wmink}):
\begin{align}
w_{\textrm{min}}(E_R^{\rm min})=\Re{\sqrt{\frac{2\delta}{\mu}}}.
\label{eq:wmin}
\end{align}
Since $w_{\textrm{min}}(E_R^{\rm min})\neq 0$ for $\delta>0$, and $w_{\textrm{min}} \rightarrow \infty$ for $E_R \rightarrow 0$ and $E_R \rightarrow \infty$ (neglecting corrections due to a finite escape velocity), in the case of endothermic scattering the rate of DM-nucleus scattering events exhibits a maximum at finite recoil energies.~This conclusion does not apply to the case $\delta \le 0$, unless the DM-nucleus scattering cross section scales with a positive power of $E_R$.

From Eqs.~(\ref{eq:wmin}) and (\ref{eq:ERmin}), one finds that for $m_\chi \gg m_T$, e.g.~$m_\chi=1$~TeV, $w_{\rm min}(E_R^{\rm min}) \simeq \Re \sqrt{2 \delta/m_T}$ and $E_R^{\rm min} \simeq \delta$.~This implies that DM-nucleus scattering is kinematically favoured in the limit of large mass number for the target nucleus, and that only direct detection experiments which record nuclear recoil energies larger than $\delta$ can be sensitive to inelastic DM models.

\subsection{Neutrino telescopes}
\label{sec:nt}
For the particles forming the Milky Way DM halo, the rate of scattering from a velocity $w$ to a velocity less than the local escape velocity at a distance $r$ from the Sun's centre, $v(r)$, is given by
\begin{align}
\OmCap =\sum_i n_i w\, \Theta(u-u_{m,i})\Theta(\EHi-E_C)\int_{\ELi}^{\EHi}\;dE_R\,\frac{d\sigma(w^2,E_R)}{dE_R},
\label{eq:omega}
\end{align}
where the index $i$ in the sum runs over the 16 most abundant elements in the Sun, namely H, $^{3}$He, $^{4}$He, $^{12}$C, $^{14}$N, $^{16}$O, $^{20}$Ne, $^{23}$Na, $^{24}$Mg, $^{27}$Al, $^{28}$Si, $^{32}$S, $^{40}$Ar, $^{40}$Ca, $^{56}$Fe, and $^{58}$Ni.~The density of the $i$-th element at a distance $r$ from the Sun's centre is denoted by $n_i(r)$ and modelled as in the {\sffamily darksusy} package~\cite{Gondolo:2004sc}.~Here $w=\sqrt{u^2+v^2(r)}$ is the DM particle velocity in the target nucleus rest frame at a distance $r$ from the centre of the Sun, while $u$ is the speed such a particle would have at infinity, which, consistently with Eq.~(\ref{eq:wmin}) must be larger than the lower bound
\begin{align}
u_{m,i}\equiv\Re{\sqrt{\frac{2\delta}{\mu_i}-v^2(r)}}\,.
\label{eq:umin}
\end{align}
Following~\cite{Gondolo:2004sc}, we compute the local escape velocity at $r$, $v(r)$, from the Sun's gravitational potential.~Finally, $\EHi$ is the maximum recoil energy, Eq.~(\ref{eq:Emax}), evaluated at the $i$-th target nucleus mass, $m_i$, i.e.~$\EHi=E_R^{{\rm max},i}\equiv E_R^{\rm max}(m_T=m_i)$.~Similarly, $\ELi=\max(E_R^{{\rm min},i}, E_C)$, where $E_R^{{\rm min},i}\equiv E_R^{\rm min}(m_T=m_i)$ and $E_C$ is the minimum energy a DM particle has to deposit in the scattering to become gravitationally bound to the Sun, i.e.~to scatter from $w$ to a velocity less than $v(r)$:
\begin{align}
E_C= \onehalf m_\chi u^2 - \delta\,.
\end{align}
Multiplying Eq.~(\ref{eq:omega}) by the rate of DM particles crossing an infinitesimal solar shell at $r$ and the time spent by each DM particle on the shell, and, finally, integrating over all radii and velocities that can contribute to the capture, one finds the known expression for the rate of DM capture in the Sun~\cite{Gould:1987ir}:
\begin{align}
C_\odot= \frac{\rho_\chi}{m_\chi} \int_0^{r_\odot} {\rm d}r \,4\pi r^2\int_0^\infty {\rm d}u \frac{f(u)}{u}w\OmCap \,,
\label{eq:C}
\end{align}
which, using Eq.~(\ref{eq:omega}), in the case of inelastic DM can be rewritten as follows
\begin{align}
C_\odot=\frac{\rho_\chi}{m_\chi}\sum_i\int^{r_{L,i}}_{0}dr 4\pi r^2n_i(r)\int_{u_{1}^i(r)}^{u_{2}^{i}(r)}du\frac{f(u)}{u}w\,
\int_{E_L^i(u,r)}^{E_H^i(u,r)}dE_R \frac{d\sigma(w^2,E_R)}{dE_R}\,.
\label{eq:Cin}
\end{align}
In the above expression, the upper and lower bound in the velocity integral arise from the Heaviside step functions in Eq.~(\ref{eq:omega}).~If the equation $E^{{\rm max},i}_R=E_C$ has two positive solutions, denoted here by $u^i_{+}$ and $u^i_{-}$, the velocity integral in Eq.~(\ref{eq:C}) must be performed between $u_1^i=u^i_{-}>u_{m,i}$ and $u_2^i=u^i_{+}$.~In this case, the Heaviside step function $\Theta(u-u_{m,i})$ in Eq.~(\ref{eq:omega}) is redundant, since integrating from $u_{m,i}$ to $u^i_{-}$ would give zero, in that $E_C>E^{{\rm max},i}_R$ in this range.~If $u^i_{\pm}$ exist, they can be found explicitly by solving $E_R^{{\rm max},i}=E_C$ for $u$.~They read as follows
\begin{align}
u_{\pm}^i = v(r) \frac{\sqrt{\,2m_\chi m_i }}{|m_\chi-m_i|}\sqrt{1+\delta\frac{m_i-m_\chi}{m_\chi^2 v^2(r)}\pm\sqrt{1+2\delta\frac{m_i-m_\chi}{m_i m_\chi v^2(r)}}}\,.
\label{eq:Cin}
\end{align}
When the equation $E^{{\rm max},i}_R=E_C$ has one positive solution only, $u^i_{+}$, the velocity integral in Eq.~(\ref{eq:Cin}) must be computed between $u_1^i=u_{m,i}$ and $u_2^i=u_{+}$.~The lower bound is in this case determined by the Heaviside step function $\Theta(u-u_{m,i})$, and $u^i_{-}$ is the only positive solution to the equation $E_R^{{\rm min},i}=E_C$.~Finally, the capture rate is zero when $E^{{\rm max},i}_R=E_C$ has no solutions.~This occurs if
\begin{align}
v(r) < v_{L,i} \equiv \Re{\sqrt{\frac{2\delta\left(m_\chi-m_i\right)}{m_\chi m_i}}}\,,
\end{align}
which implies an upper bound for the radial integral given by $r_{L,i}=r(v_{L,i})$, where $v\rightarrow r(v)$ is the inverse of the monotonic function $r\rightarrow v(r)$.~Notice that $r_{L,i}=0$ for $v_{L,i} > v(0)$, and $r_{L,i}=r_\odot$ for $v_{L,i} < v(r_\odot)$.~From the practical point of view, it is convenient to perform the velocity integral in Eq.~(\ref{eq:Cin}) by assuming $u_1^i$ equal to $u_{m,i}$, and setting $C_\odot=0$ whenever $E_C>E^{{\rm max},i}_R$.

Assuming equilibrium between capture and annihilation, the rate of DM annihilations in the Sun is given by $\Gamma_a=C_\odot/2$.~While this assumption applies to the case of DM capture via elastic DM-nucleus scattering, it is not generically fulfilled by inelastic DM models where contributions to the elastic scattering cross section are smaller than about $10^{-48}$~cm$^2$~\cite{Menon:2009qj}.~In this latter case, one has to multiply $\Gamma_a$ by a correction factor, $\eta(m_\chi,\delta)$, which accounts for the lack of DM thermalisation in the Sun and the associated changes in the radial distribution of captured DM particles.~The correction factor $\eta(m_\chi,\delta)$ has been computed in~\cite{Blennow:2018xwu} for $0\le\delta\le200$~keV, $100\le m_\chi\le 500$~GeV and assuming a fully inelastic spin-independent cross section in the range $10^{-42}$-$10^{-45}$~cm$^2$.~Presenting our constraints on inelastic DM from IceCube we will need values for $\eta$ outside this range of $\delta$ and $m_\chi$.~While a Monte Carlo simulation of the thermalisation process would be the best approach to estimate $\eta$ in the large $\delta$ and $m_\chi$ limits, such a detailed calculation goes beyond the scope of the present article.~Instead, we will perform a 2-dimensional linear extrapolation in the plane spanned by $\ln \delta$ and $\ln m_\chi$ of the function $\eta(m_\chi,\delta)$ obtained in~\cite{Blennow:2018xwu} for an inelastic DM-nucleon scattering cross section of $10^{-45}$~cm$^2$.~This estimate for $\eta$ is expected to be conservative, i.e.~smaller than its actual value.~One of the reasons is that it relies on an inelastic DM-nucleon scattering cross section of $10^{-45}$~cm$^2$, which is smaller than the cross section values IceCube can constrain, and, in general, a larger inelastic DM-nucleon scattering cross section would lead to a larger $\eta$~\cite{Blennow:2018xwu}.~A second reason is that it assumes that the elastic cross section for DM-nucleon scattering is exactly zero, which is in general not true in concrete models for inelastic DM (see for example~\cite{Bramante:2016rdh}).

\begin{figure}[t]
    \centering \includegraphics[trim={3cm, 0cm, 4cm, 0cm},width=0.5\textwidth]{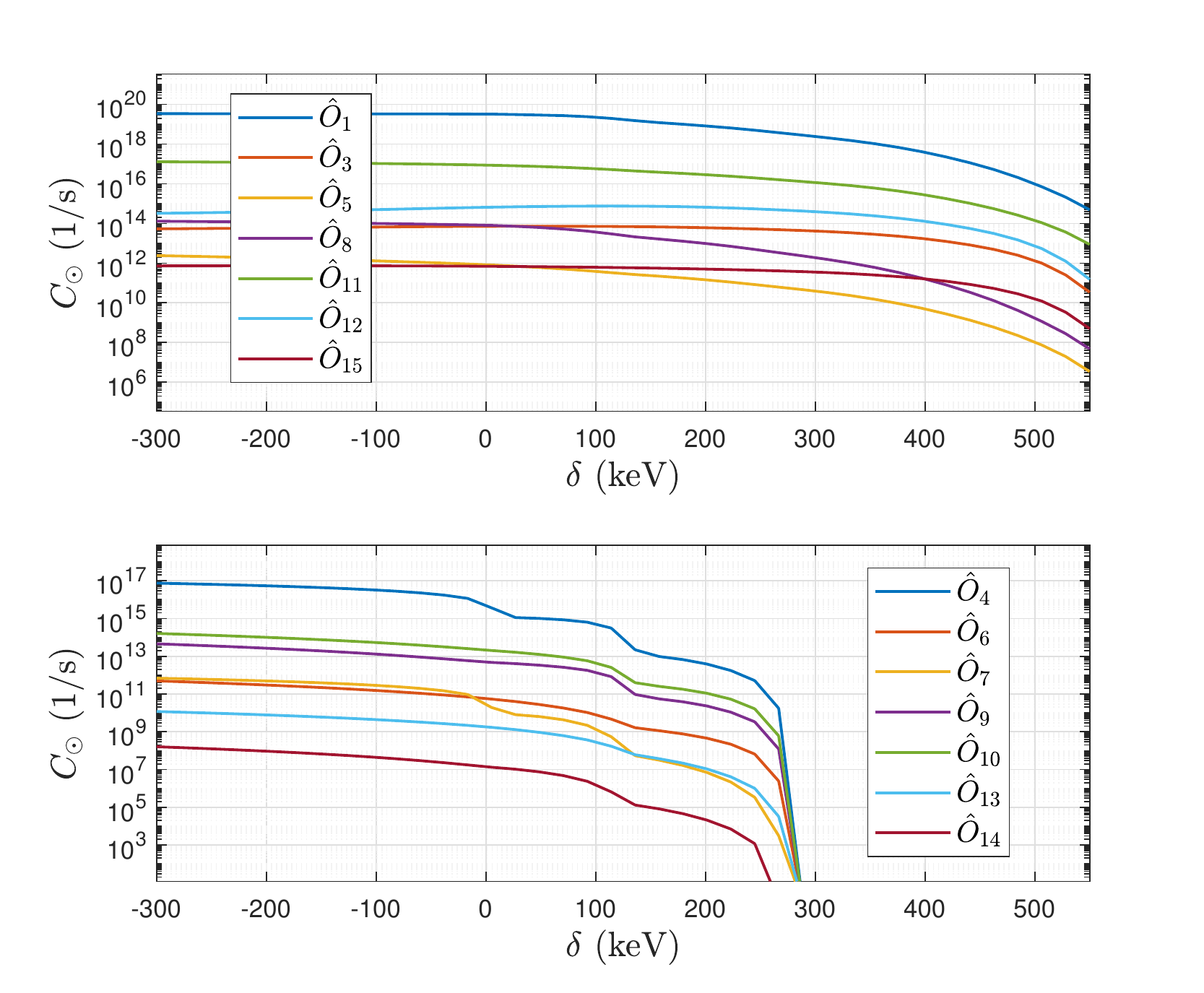}
    \caption{Capture rate as a function of the mass splitting for the isoscalar operators in Tab.~\ref{tab:operators}.~We set DM particle spin and mass to 1/2 and 1 TeV, respectively.~For the coupling constants, we assume $c_j^0=10^{-3}$ (246.2 GeV)$^{-2}$.}
    \label{fig:Cisos}
\end{figure} 

\begin{figure}[t]
    \centering \includegraphics[trim={3cm, 0cm, 4cm, 0cm},width=0.5\textwidth]{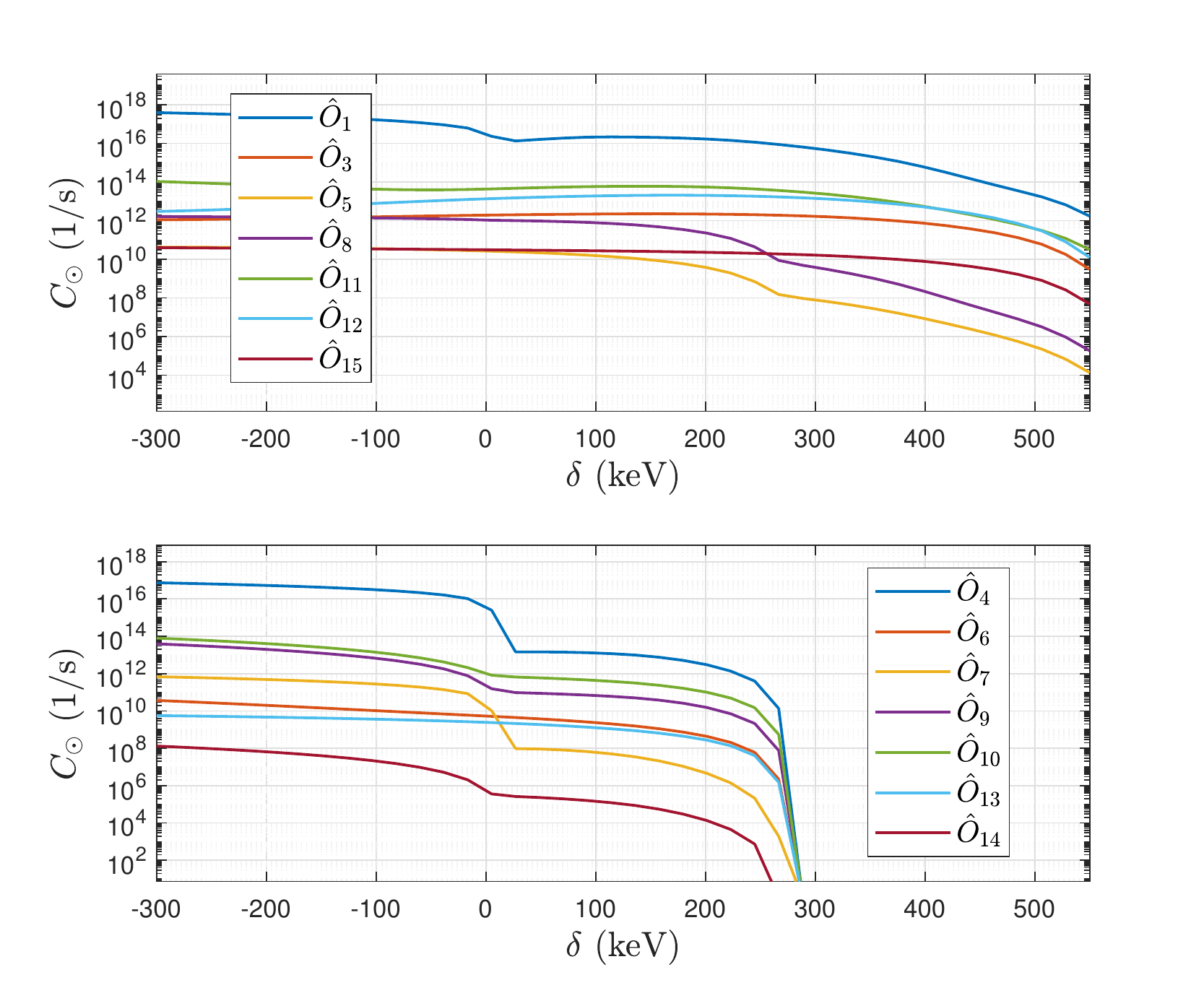}
    \caption{Same as Fig.~\ref{fig:Cisos}, but now for the isovector component of the operators in Tab.~\ref{tab:operators}.~We set $c_j^1=10^{-3}$ (246.2 GeV)$^{-2}$.}
    \label{fig:Cisov}
\end{figure}

The differential neutrino flux from DM annihilation in the Sun depends linearly on $\Gamma_a$ and is given by~\cite{Jungman:1995df}
\begin{equation}
    \frac{{\rm d} \Phi_\nu}{{\rm d} E_\nu} =
    \frac{\Gamma_a}{4\pi D^2}\sum_f B^f_\chi \frac{{\rm d} N^f_\nu}{{\rm d} E_\nu}\,.
\label{eq:nuflux}
\end{equation}
In Eq.~(\ref{eq:nuflux}), $B^f_\chi$ is the branching ratio for DM pair annihilation into the final state $f$, ${\rm d} N^f_\nu / {\rm d} E_\nu$ is the neutrino energy spectrum at the detector from the decay of Standard Model particles into the final state $f$, $E_\nu$ is the neutrino energy and $D$ is the detector's distance to the Sun's centre.~The associated DM-induced differential muon flux at neutrino telescopes is given by
\begin{equation}
\frac{{\rm d}\Phi_{\mu}}{{\rm d} E_\mu} = N_T \int_{E_\mu^{\rm th}}^{\infty} {\rm d} E_\nu
\int_0^{\infty} {\rm d} \lambda \int_{E_\mu}^{E_\nu} {\rm d} E_\mu^{\prime} \,\mathcal{P}(E_\mu,E_\mu^{\prime};\lambda)\, \frac{{\rm d}\sigma_{{\rm CC} }(E_\nu,E_\mu^{\prime})}{{\rm d} E_{\mu}^{\prime}}  \,\frac{{\rm d \Phi_\nu}}{{\rm d} E_\nu} \,,
\label{eq:Phi}
\end{equation}
where $N_T$ is the number of nucleons per cubic centimetre, $E_\mu^{\rm th}$ is the detector energy threshold, $\lambda$ is the muon range, $\mathcal{P}(E_\mu,E_\mu^{\prime};\lambda)$ is the probability for a muon of initial energy $E_{\mu}^{\prime}$ to be detected with a final energy $E_\mu$ after traveling a distance $\lambda$ inside the detector, and ${\rm d}\sigma_{{\rm CC}}/{\rm d} E_\mu^{\prime}$ is the weak differential cross section for production of a muon of energy $E_\mu^{\prime}$.~In our analysis, we evaluate Eq.~(\ref{eq:Phi}) using neutrino yields generated by {\sffamily WimpSim}~\cite{Blennow:2007tw}, and tabulated in {\sffamily darksusy}~\cite{Gondolo:2004sc}.

\section{Constraining the large mass splitting limit}
\label{sec:results}
In this section we evaluate our expression for the capture rate, Eq.~(\ref{eq:Cin}), under different assumptions (Sec.~\ref{sec:capture}) and use these results to set constraints on inelastic DM from the IceCube neutrino telescope (Sec.~\ref{sec:constraints}).~Such constraints will be compared with those from direct detection experiments.~In the analysis, emphasis will be placed on the large mass splitting limit.

\subsection{Generalised solar capture rate}
\label{sec:capture}
Fig.~\ref{fig:Cisos} shows the capture rate $C_\odot$ in Eq.~(\ref{eq:Cin}) for the isoscalar component of the operators $\hat{\mathcal{O}}_j$, $j=1,3,\dots,15$ in Tab.~\ref{tab:operators} as a function of the mass splitting parameter $\delta$.~Here we focus on spin 1/2 DM and $m_\chi=1$~TeV, and do not consider the operators $\hat{\mathcal{O}}_{17}$ and $\hat{\mathcal{O}}_{18}$ which can only arise for spin 1 DM.~The operators in Fig.~\ref{fig:Cisos} naturally divide into two families of 7 operators each.~The first family consists of the operators that generate scattering cross sections which are not zero for spin 0 nuclei.~In this case, DM can effectively be captured in the Sun for mass splittings larger than 500 keV due to the relatively large mass numbers of the spin 0 nuclei $^{56}$Fe and $^{58}$Ni (see discussion at the end of Sec.~\ref{sec:kinematics}).~Operators belonging to this family are $\hat{\mathcal{O}}_1$, $\hat{\mathcal{O}}_3$, $\hat{\mathcal{O}}_5$, $\hat{\mathcal{O}}_8$, $\hat{\mathcal{O}}_{11}$, $\hat{\mathcal{O}}_{12}$, and $\hat{\mathcal{O}}_{15}$.~The second family consists of operators that generate cross sections which are zero for spin 0 nuclei.~The fact that the heaviest nucleus in the Sun with spin different from zero (and sufficiently abundant) is $^{27}$Al explains the sharp decrease in the capture rate found for $\delta$ below 300 keV in the bottom panel of Fig.~\ref{fig:Cisos}.~Operators belonging to this second family are $\hat{\mathcal{O}}_4$, $\hat{\mathcal{O}}_6$, $\hat{\mathcal{O}}_7$, $\hat{\mathcal{O}}_9$, $\hat{\mathcal{O}}_{10}$, $\hat{\mathcal{O}}_{13}$, and $\hat{\mathcal{O}}_{14}$.~We find similar conclusions for the isovector component of the operators in Tab.~\ref{tab:operators}.~Results for the isovector couplings are illustrated in Fig.~\ref{fig:Cisov}.

\begin{table}[t]
\centering
\begin{tabular}{|c||c|c|c|c|c|c|c|c|c|c|c|c|}\hline
\diagbox{$\hO_j$}{$\delta$}  &-300 &-213 &-147 &-60 &5 &92 &158 &245 &310 &397 &463 &550\\\hline\hline
$\hO_1$ &$^4$He &O &O &O &O &O &Fe &Fe &Fe &Fe &Fe &Fe\\\hline
$\hO_{3}$ &Si &Si &Si &Fe &Fe &Fe &Fe &Fe &Fe &Fe &Fe &Fe\\\hline
$\hO_{4}$ &H &H &H &H &H &N &Al &Al &-- &-- &-- &--\\\hline
$\hO_{5}$ &N &N &N &N &N &Fe &Fe &Fe &Fe &Fe &Fe &Fe\\\hline
$\hO_{6}$ &N &N &N &N &N &N &Al &Al &-- &-- &-- &--\\\hline
$\hO_{7}$ &H &H &H &H &H &N &Al &Al &-- &-- &-- &--\\\hline
$\hO_{8}$ &N &N &N &N &N &Fe &Fe &Fe &Fe &Fe &Fe &Fe\\\hline
$\hO_{9}$ &H &H &H &N &N &N &Al &Al &-- &-- &-- &--\\\hline
$\hO_{10}$ &N &N &N &N &N &N &Al &Al &-- &-- &-- &--\\\hline
$\hO_{11}$ &O &O &O &O &O &Fe &Fe &Fe &Fe &Fe &Fe &Fe\\\hline
$\hO_{12}$ &Si &Si &Si &Fe &Fe &Fe &Fe &Fe &Fe &Fe &Fe &Fe\\\hline
$\hO_{13}$ &N &N &N &N &N &N &Al &Al &-- &-- &-- &--\\\hline
$\hO_{14}$ &H &H &H &N &N &N &Al &Al &-- &-- &-- &--\\\hline
$\hO_{15}$ &Si &Fe &Fe &Fe &Fe &Fe &Fe &Fe &Fe &Fe &Fe &Fe\\\hline
\end{tabular}
\caption{Elements in the Sun with the largest capture rate for the isoscalar operators in Tab.~\ref{tab:operators} as a function of $\delta$, in keV, and for $m_\chi=1$~TeV.~When capture is kinematically not allowed, that entry is filled in with a dash.}
\label{tab:capis}
\end{table}

\begin{table}[t]
\centering
\begin{tabular}{|c||c|c|c|c|c|c|c|c|c|c|c|c|}\hline
\diagbox{$\hO_j$}{$\delta$} &-300 &-213 &-147 &-60 &5 &92 &158 &245 &310 &397 &463 &550\\\hline\hline
$\hO_1$ &H &H &H &H &H &Fe &Fe &Fe &Fe &Fe &Fe &Ni\\\hline
$\hO_{3}$ &Fe &Fe &Fe &Fe &Fe &Fe &Fe &Fe &Fe &Fe &Fe &Fe\\\hline
$\hO_{4}$ &H &H &H &H &H &Al &Al &Al &-- &-- &-- &--\\\hline
$\hO_{5}$ &Al &Al &Al &Al &Al &Al &Al &Al &Fe &Fe &Fe &Ni\\\hline
$\hO_{6}$ &Al &Al &Al &Al &Al &Al &Al &Al &-- &-- &-- &--\\\hline
$\hO_{7}$ &H &H &H &H &H &Al &Al &Al &-- &-- &-- &--\\\hline
$\hO_{8}$ &Al &Al &Al &Al &Al &Al &Al &Al &Fe &Fe &Fe &Ni\\\hline
$\hO_{9}$ &H &H &H &H &Al &Al &Al &Al &-- &-- &-- &--\\\hline
$\hO_{10}$ &H &H &H &H &Al &Al &Al &Al &-- &-- &-- &--\\\hline
$\hO_{11}$ &H &H &H &Fe &Fe &Fe &Fe &Fe &Fe &Fe &Fe &Ni\\\hline
$\hO_{12}$ &Fe &Fe &Fe &Fe &Fe &Fe &Fe &Fe &Fe &Fe &Fe &Fe\\\hline
$\hO_{13}$ &Al &Al &Al &Al &Al &Al &Al &Al &-- &-- &-- &--\\\hline
$\hO_{14}$ &H &H &H &H &Al &Al &Al &Al &-- &-- &-- &--\\\hline
$\hO_{15}$ &Fe &Fe &Fe &Fe &Fe &Fe &Fe &Fe &Fe &Fe &Fe &Fe\\\hline
\end{tabular}
\caption{Same as Tab.~\ref{tab:capiv}, but now for the isovector interactions.}
\label{tab:capiv}
\end{table}

Depending on the interaction type, and on the value of the mass splitting parameter, different elements in the Sun give the largest contribution to the capture rate.~For a sample of (positive and negative) values of $\delta$, these elements are listed in Tab.~\ref{tab:capis} for the isoscalar interactions, and in Tab.~\ref{tab:capiv} for the isovector interactions.~In both tables, we assume spin 1/2 DM and $m_\chi=1$~TeV.

\begin{figure}
    \centering \includegraphics[trim={3cm, 0cm, 4cm, 0cm},width=0.7\textwidth]{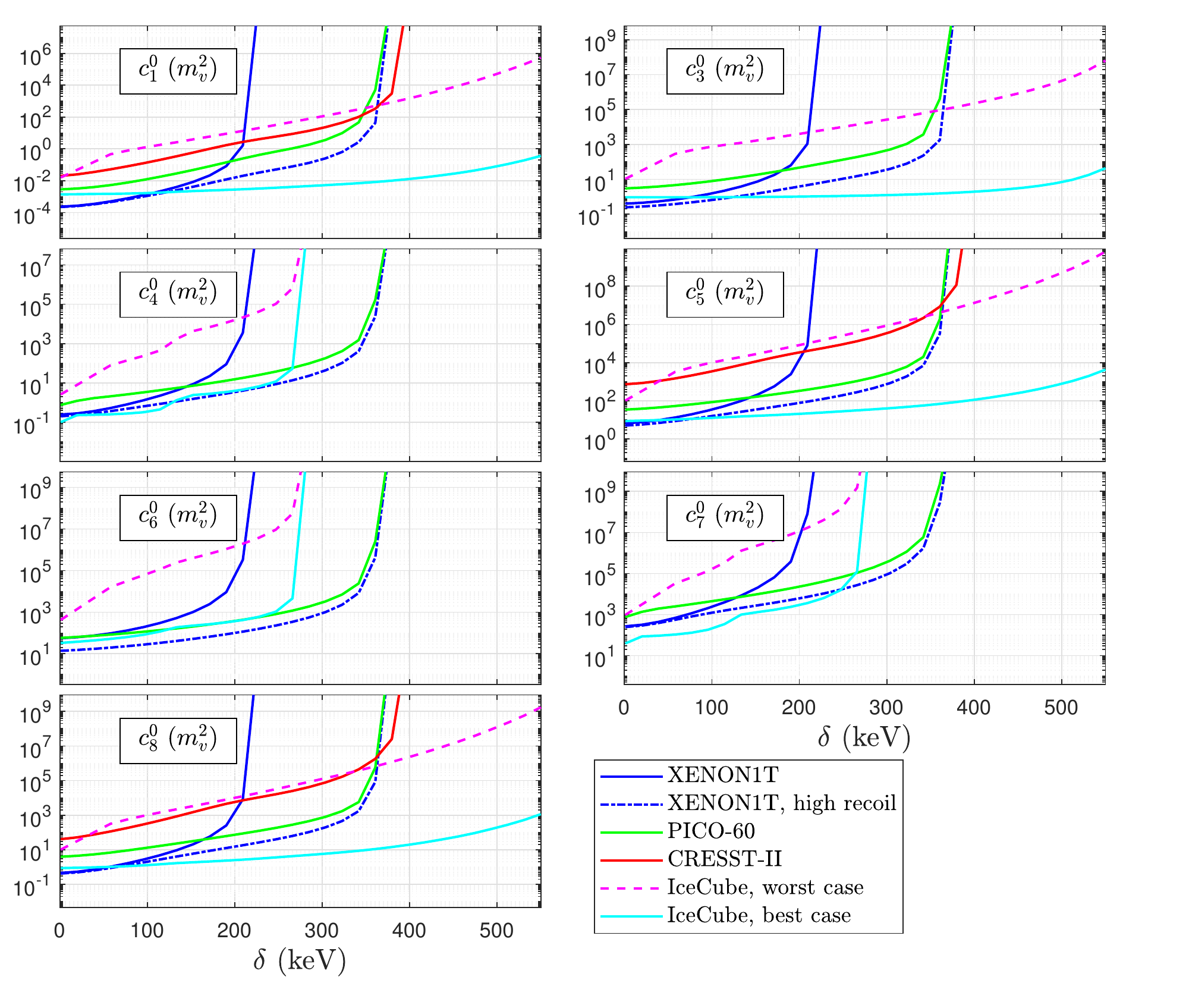}
    \caption{Constraints on the isoscalar coupling constants for operators $\hat{O}_1$ to $\hat{O}_8$ for 1 TeV DM at a 90\% confidence level from XENON1T, a hypothetical high-recoil analysis of XENON1T, PICO-60, CRESST-II and IceCube.}
    \label{fig:isosc1}
\end{figure}

\begin{figure}
    \centering \includegraphics[trim={3cm, 0cm, 4cm, 0cm},width=0.7\textwidth]{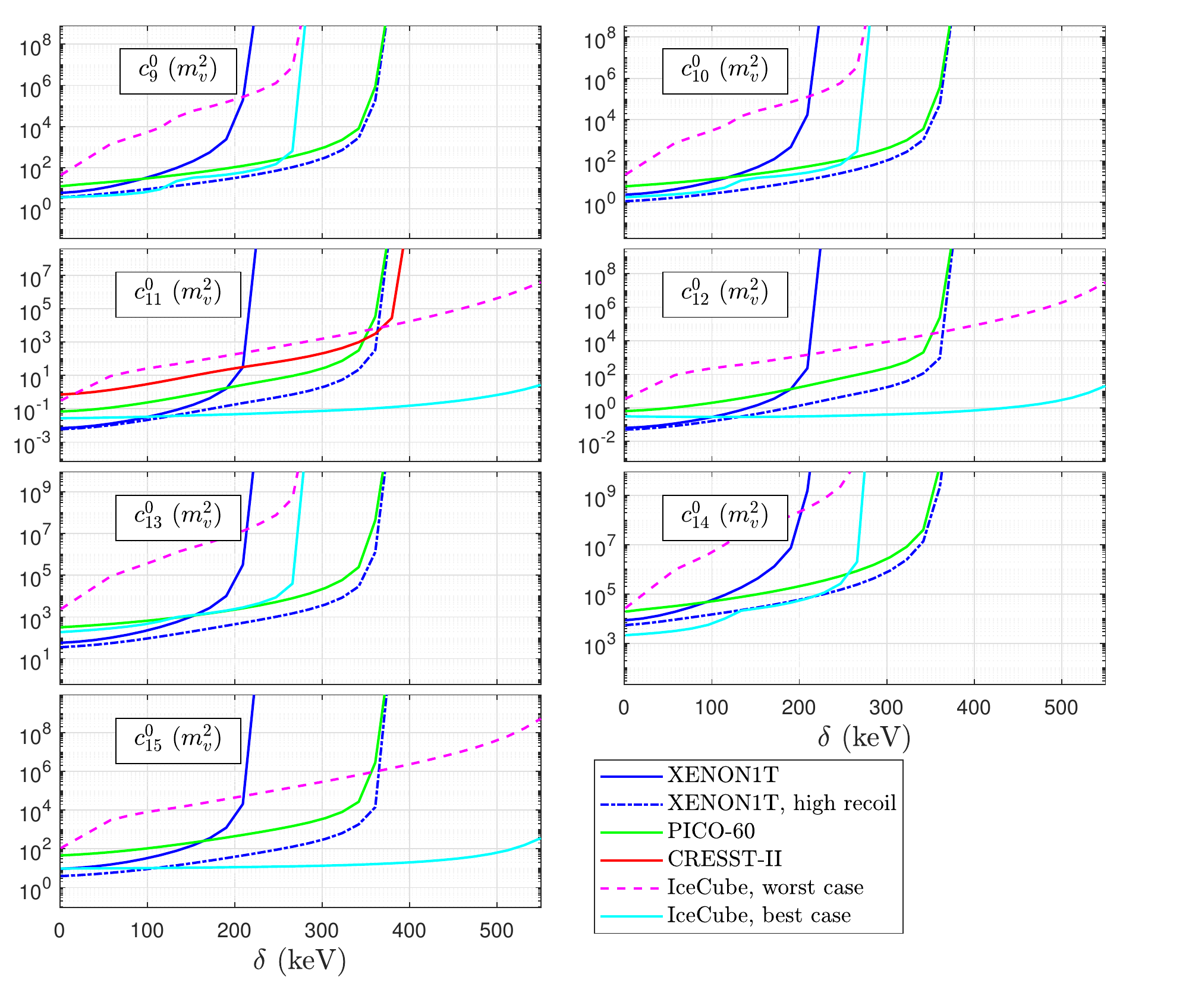}
    \caption{Constraints on the isoscalar coupling constants for operators $\hat{O}_9$ to $\hat{O}_{15}$ for 1 TeV DM at a 90\% confidence level from XENON1T, a hypothetical high-recoil analysis of XENON1T, PICO-60, CRESST-II and IceCube.}
    \label{fig:isosc2}
\end{figure}

\begin{figure}
    \centering \includegraphics[trim={3cm, 0cm, 4cm, 0cm},width=0.7\textwidth]{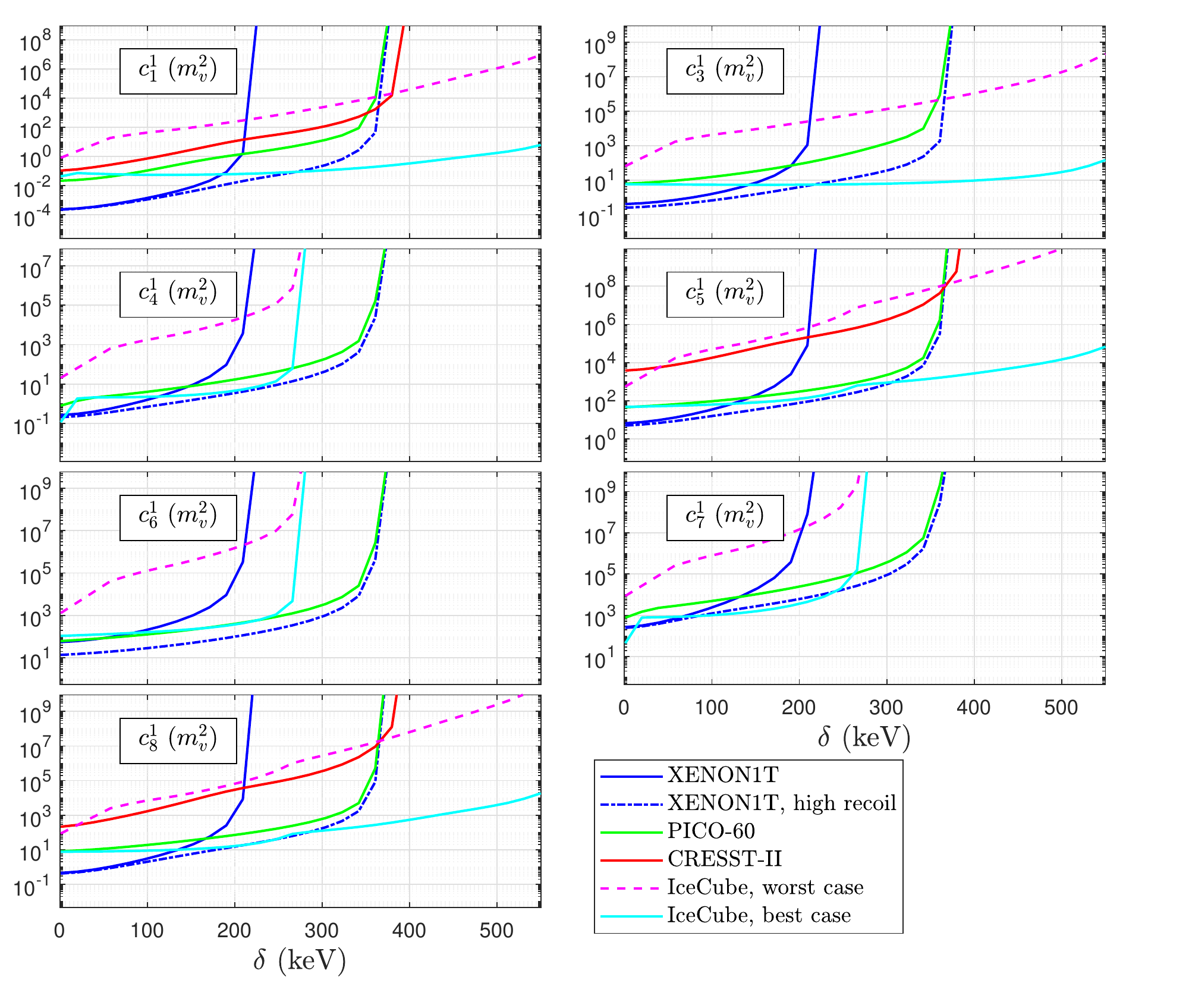}
    \caption{Constraints on the isovector coupling constants for operators $\hat{O}_1$ to $\hat{O}_8$ for 1 TeV DM at a 90\% confidence level from XENON1T, a hypothetical high-recoil analysis of XENON1T, PICO-60, CRESST-II and IceCube.}
    \label{fig:isovc1}
\end{figure}

\begin{figure}
    \centering \includegraphics[trim={3cm, 0cm, 4cm, 0cm},width=0.7\textwidth]{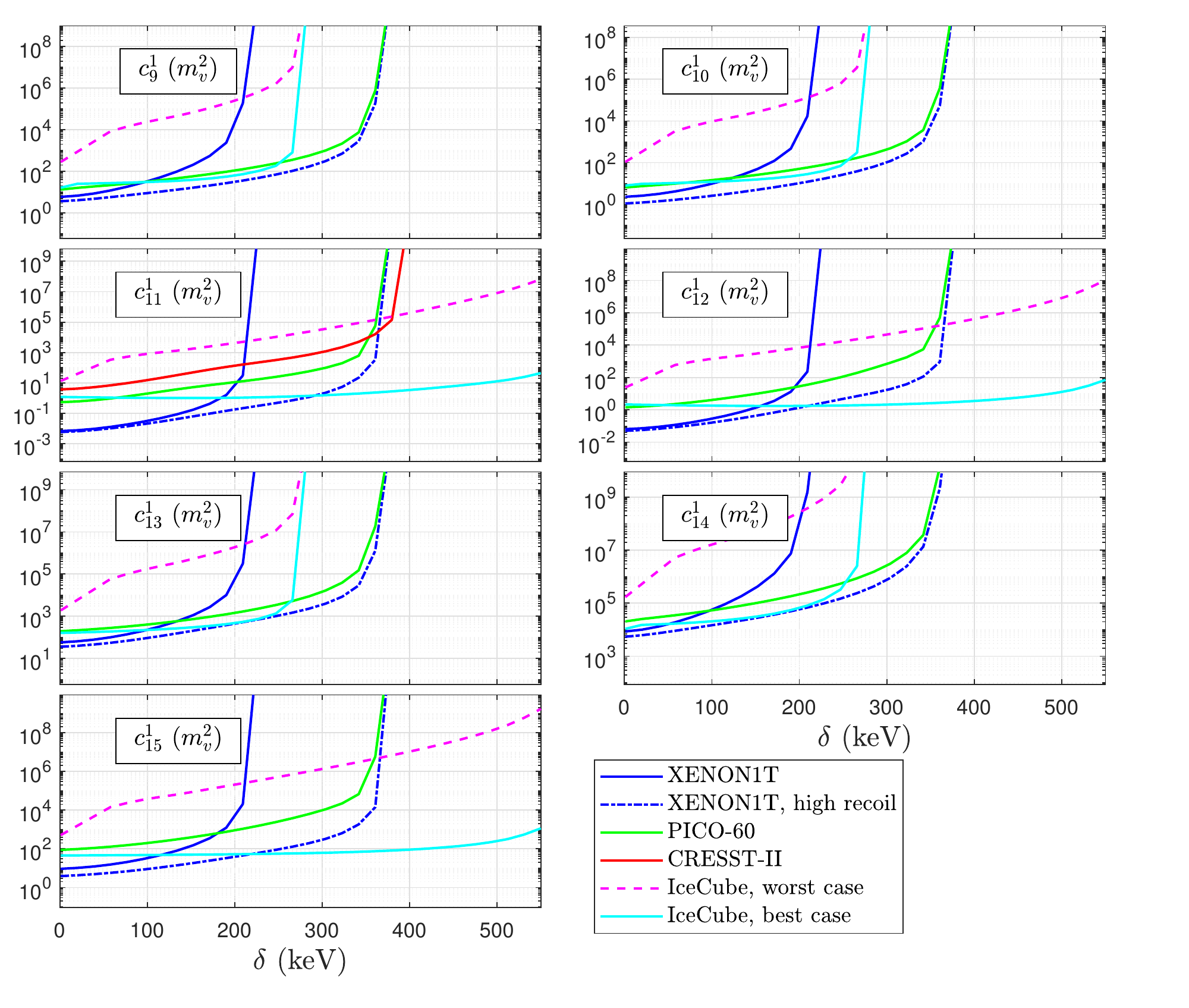}
    \caption{Constraints on the isovector coupling constants for operators $\hat{O}_9$ to $\hat{O}_{15}$ for 1 TeV DM at a 90\% confidence level from XENON1T, a hypothetical high-recoil analysis of XENON1T, PICO-60, CRESST-II and IceCube.}
    \label{fig:isovc2}
\end{figure}

\subsection{Exclusion limits from IceCube (and direct detection)}
\label{sec:constraints}

In this subsection, we use Eqs.~(\ref{eq:rate}) and (\ref{eq:Phi}), and the results in Sec.~\ref{sec:capture}, to set 90\% C.L. exclusion limits on the coupling constants for isoscalar and isovector DM-nucleon interactions, $c_j^0$ and $c_j^1$, respectively.~Limits are computed using results from the XENON1T~\cite{Aprile:2018dbl}, PICO-60~\cite{Amole:2015pla} and CRESST-II~\cite{Angloher:2015ewa} direct detection experiments, as well as data from the IceCube neutrino telescope~\cite{Aartsen:2016zhm}.~Computing the expected number of signal events at XENON1T, PICO-60 and CRESST-II, we assume the following exposures and nuclear recoil energy intervals:~1300$\times$279 kg$\times$day and 5-40 keV for XENON1T, 52 kg$\times$day and 10-1000 keV for PICO-60, and 1300 kg$\times$day and 0.3-120 keV for CRESST-II.~We also implement a hypothetical high-recoil energy analysis of XENON1T, with energy range 5-240 keV~\cite{Aprile:2017aas}.~We intentionally do not use the more recent PICO results based on a C$_3$F$_8$ detector~\cite{Amole:2017dex}, since C$_3$FI (used in~\cite{Amole:2015pla}) is a better target material to set constraints on inelastic DM, because of the large mass number of iodine (see discussion at the end of Sec.~\ref{sec:kinematics}).~Similarly, we do not use CRESST-III results~\cite{Petricca:2017zdp}, since CRESST-II has a larger exposure.~Exclusion limits from IceCube are computed by requiring that the total muon flux obtained from Eq.~(\ref{eq:Phi}) is smaller than the corresponding upper bound in Tab.~IV of~\cite{Aartsen:2016zhm}.~For definiteness, here we focus on two limiting cases:~1) A first one where DM thermalises in the Sun (e.g.~via sub-leading elastic interactions) and annihilates into a $\tau^-\tau^+$ pair;~2) A second one where DM does not thermalise, $\Gamma_a$ must be multiplied by the correction factor $\eta$, and the annihilation channel is $b\bar{b}$.~We refer to the first and second scenario as ``best case'' and ``worst case'', respectively.~Exclusion limits from direct detection experiments are computed by requiring that the total number of signal events is less than 2.3 in PICO-60, and  is less than 7.99 in XENON1T and CRESST-II.~This procedure corresponds to assuming Poisson statistics for the observed number of nuclear recoil events in the signal region, which we set to 0 in the case of PICO-60, and to 4 in the case of XENON1T and CRESST-II.~Finally, exclusion limits are presented as a function of the mass splitting parameter $\delta$, focusing on a DM candidate of mass 1 TeV and spin 1/2, and on the range $0 \le \delta \le 550$~keV.

Figs.~\ref{fig:isosc1} and \ref{fig:isosc2} (\ref{fig:isovc1} and \ref{fig:isovc2}) show our exclusion limits on the coupling constants $c_j^0$ ($c_j^1$).~In the case of CRESST, we only compute the exclusion limits for the operators $\hat{\mathcal{O}}_1$, $\hat{\mathcal{O}}_5$, $\hat{\mathcal{O}}_{8}$ and $\hat{\mathcal{O}}_{11}$.~The reason is that for $m_\chi=1$~TeV tungsten gives the largest contribution to the event rate at CRESST, and nuclear response functions for W are not available for operators different from the ones listed above.~For these operators, the use of Helm form factors is a good first approximation.~In order to describe the results in Figs.~\ref{fig:isosc1}, \ref{fig:isosc2}, \ref{fig:isovc1} and \ref{fig:isovc2}, let us examine the case of isoscalar DM-nucleon interactions of type $\hat{\mathcal{O}}_1$ in some detail.~This example illustrates one of the main results of this article:~in the large mass splitting limit, neutrino telescopes place the most stringent limits on the coupling constants of all spin-independent DM-nucleus interactions.~In the specific case under consideration, XENON1T places the strongest direct detection exclusion limits on $c_1^0$ for $\delta\le 210$~keV.~However, for $\delta>210$~keV, a large fraction of the predicted DM-nucleus scattering events lie outside the XENON1T signal region, and PICO places the strongest direct detection limits on $c_1^0$, since it records data up to $E_R=1000$~keV.~On the other hand, above $\delta=320$~keV, it is CRESST that sets the strongest direct detection limits on $c_1^0$ because of the large mass number of tungsten.~Our results also show that a hypothetical high-recoil energy analysis of XENON1T data would give the most stringent direct detection limits on $c_1^0$ for $\delta<360$~keV, but would not improve CRESST constraints above $\delta=360$~keV.~Furthermore, we find that for the interaction operator $\hat{\mathcal{O}}_1$ (as well as for the other momentum transfer-independent operators) the advantages of a high-recoil energy analysis are only significant for $\delta$ larger than around $100$~keV.~Finally, we find that IceCube sets the strongest exclusion limits on the coupling $c_1^0$ in a wide range of values for the mass splitting parameter.~Specifically, for the ``worst case'' scenario, IceCube gives the most stringent limits for $\delta>360$~keV, and for the ``best case'' scenario, IceCube places the strongest limits already for $\delta>150$~keV.

The exclusion limits on inelastic DM that we obtain from IceCube have a significant impact on DM models.~As an example of a specific model constrained by our results, let us mention a nearly pure higgsino DM.~Such a DM candidate is expected to have a sufficiently large elastic DM-nucleus scattering cross section to thermalise in the Sun, which implies an unsuppressed muon flux~\cite{Fox:2014moa}.~Once constraints on the present DM cosmological density have been imposed, the expected value for $c_1^0$ is in this case $c_1^0 \simeq 0.35$ (246.2 GeV)$^{-2}$~\cite{Bramante:2016rdh}.~Therefore, this model is excluded by the limits presented here for mass splittings up to 530 keV.~On the other hand, previous analyses which only accounted for direct detection results could not exclude a pure higgsino as a DM candidate for values of the mass splitting parameter larger than $\delta=220$~keV.~A high-recoil energy analysis of XENON1T data would only rule out this model for $\delta<320$~keV.

Inspection of the results in Figs.~\ref{fig:isosc1}, \ref{fig:isosc2}, \ref{fig:isovc1} and \ref{fig:isovc2} shows that interaction operators can be divided into two families (as in the previous subsection).~The first family consists of the operators that generate DM-nucleus scattering cross sections that are not zero for spin 0 nuclei, namely $\hat{\mathcal{O}}_1$, $\hat{\mathcal{O}}_3$, $\hat{\mathcal{O}}_5$, $\hat{\mathcal{O}}_8$, $\hat{\mathcal{O}}_{11}$, $\hat{\mathcal{O}}_{12}$ and $\hat{\mathcal{O}}_{15}$.~For these operators, the most stringent exclusion limits on $c_j^0$ and $c_j^1$ come from direct detection experiments in the small mass splitting limit, and from neutrino telescopes for large mass splittings.~This change in the hierarchy of constraints is due to kinematical reasons.~Indeed, while even for large $\delta$ DM can still be captured in the Sun (as long as the kinematic constraints in Sec.~\ref{sec:inelastic} are fulfilled), for sufficiently large $\delta$ most of the DM particles in the Milky Way halo move with a speed smaller than $w_{\rm min}$ in Eq.~(\ref{eq:wmin}), and cannot induce observable recoils at direct detection experiments.~The second family consists of the operators that generate DM-nucleus scattering cross sections which are zero for spin 0 nuclei, namely $\hat{\mathcal{O}}_4$, $\hat{\mathcal{O}}_6$, $\hat{\mathcal{O}}_7$, $\hat{\mathcal{O}}_9$, $\hat{\mathcal{O}}_{10}$, $\hat{\mathcal{O}}_{13}$ and $\hat{\mathcal{O}}_{14}$.~For these operators, the capture of DM in the Sun can only occur via scattering on H, $^{3}$He, $^{14}$N, $^{23}$Na and $^{27}$Al, since other elements have spin 0.~Based on the results of Sec.~\ref{sec:capture}, this implies that IceCube is not sensitive to such interactions as long as $\delta$ is larger than 285 keV.~Indeed, for $\delta>285$~keV, $^{27}$Al, the heaviest element in the Sun with spin different from zero, is no longer able to capture DM.

Let us now focus on the comparison of our exclusion limits on the isoscalar coupling constants with the corresponding exclusion limits on the isovector couplings.~We find that limits on the isovector couplings from IceCube are less competitive than the ones on their isoscalar counterparts.~The reason is that spin-idependent isovector interactions probe the proton-neutron difference within nuclei in the Sun, and the latter is comparatively small.~For many of the most abundant nuclei in the Sun this difference is 0, and the maximum difference is 4 (for iron and argon).~In contrast, all xenon (tungsten) isotopes have a difference of at least 16 (32).

We conclude this section by briefly commenting on how our results would qualitatively change if we considered a different value for the DM particle mass.~We have checked numerically that for $m_\chi \neq 1$~TeV our results remain qualitatively unchanged.~For example, for $m_\chi \neq 1$~TeV, and assuming DM-nucleon interactions of type  ${\mathcal{O}}_1$,  ${\mathcal{O}}_3$, ${\mathcal{O}}_5$, $\hat{\mathcal{O}}_8$, $\hat{\mathcal{O}}_{11}$, $\hat{\mathcal{O}}_{12}$ or $\hat{\mathcal{O}}_{15}$, neutrino telescopes still set exclusion limits on isoscalar and isovector coupling constants that are more stringent than the ones from direct detection experiments for sufficiently large values of $\delta$.

\section{Conclusion}
\label{sec:conclusion}

We have studied the capture and subsequent annihilation of inelastic DM in the Sun, placing constraints on the DM-nucleon scattering cross section (which is quadratic in $c_j^0$ and $c_j^1$) from the null result of IceCube.~The cross section for inelastic DM-nucleon scattering has been calculated within an extension of the effective theory of DM-nucleon interactions which applies to the case of inelastic DM.~We have explicitly performed our calculations assuming a DM particle mass of 1 TeV, but our conclusions qualitatively also apply to DM particle candidates with different masses.~We find that for values of the mass splitting parameter larger than about 200 keV neutrino telescopes place limits on the DM-nucleon scattering cross section which are stronger than the ones from current DM direct detection experiments.~The exact mass splitting value depends on whether DM thermalises in the Sun or not.~This result applies to all DM-nucleon interactions that generate DM-nucleus scattering cross sections which do not depend on the nuclear spin, including the ``canonical'' spin-independent interaction, i.e.~operator $\hat{\mathcal{O}}_1$ in Tab.~\ref{tab:operators}.~Indeed, for these interactions IceCube exclusion limits on the corresponding coupling constants remain relatively flat up to mass splittings of about 300 keV.~Furthermore, we find that exclusion limits from IceCube on the coupling constants of this family of interactions are more stringent than the ones from a (hypothetical) reanalysis of XENON1T data based on an extended signal region in nuclear recoil energy.~Our results should be taken into account in the analysis of neutrino telescope data, and in global statistical analysis of inelastic DM models.

\acknowledgments It is a great pleasure to thank Anton B\"ackstr\"om, Anastasia Danopoulou, K\aa re Fridell, Martin B.~Krauss and Vanessa Zema for useful and interesting discussion on dark matter direct detection and capture in the Sun.~This work was supported by the Knut and Alice Wal- lenberg Foundation and is partly performed within the Swedish Consortium for Dark Matter Direct Detection (SweDCube).


\providecommand{\href}[2]{#2}\begingroup\raggedright\endgroup

\end{document}